\documentclass[12pt,a4paper,final]{iopart}
\usepackage{iopams}  
\usepackage{graphicx}
\usepackage[breaklinks=true,colorlinks=true,linkcolor=blue,urlcolor=blue,citecolor=blue]{hyperref}

\usepackage{appendix}

\usepackage{bm}

\begin{document}

\title[Optically tunable spin transport on the surface of a topological insulator]{Optically tunable spin transport on the surface of a topological insulator}

\author{D Yudin$^{1,2}$, O V Kibis$^{3,2,4}$ and I A Shelykh$^{1,2,4}$}
\address{$^1$ITMO University, Saint Petersburg 197101, Russia}
\address{$^2$Division of Physics and Applied Physics, Nanyang Technological University 637371, Singapore}
\address{$^3$Department of Applied and Theoretical Physics, Novosibirsk State Technical University, Novosibirsk 630073, Russia}
\address{$^4$Science Institute, University of Iceland IS-107, Reykjavik, Iceland}
\ead{dimafizmath@ya.ru}

\begin{abstract}
The emerging field of spinoptronics has a potential to supersede the functionality of modern electronics, while a proper description of strong light-matter coupling pose the most intriguing questions from both fundamental scientific and technological perspectives. In this paper we address a highly relevant issue for such a development. We theoretically explore spin dynamics on the surface of a 3D topological insulator (TI) irradiated with an off-resonant high-frequency electromagnetic wave. The strong coupling between electrons and the electromagnetic wave drastically modifies the spin properties of TI. The effects of irradiation are shown to result in anisotropy of electron energy spectrum near the Dirac point and suppression of spin current and are investigated in detail in this work.
\end{abstract}

\vspace{2pc}

\maketitle

\section{Introduction}

Since the discovery of topological insulators (TI) \cite{Kane2007,Hasan2008,Hasan2009,Hasan2014,Kane,Zhang} followed by unprecedented research activity in the field there has been a tremendous interest in studying their unique spin properties \cite{Murakami,Mclver,STT,Saitoh,Ando,Li,Liu,Black}. The phenomenon which is called topological protection makes it harder to destroy the quantum phase coherence of conducting electrons propagating at the surface. In the meanwhile, theoretical description of light-matter coupling in TIs still remains a challenge, representing thus an exciting part of both experimental and theoretical condensed matter \cite{Cayssol2013}. In most cases theoretical description is restricted to the regime of weak light-matter coupling. If this is the case, the electron-field interaction can be treated as a small perturbation which does not change an electron energy spectrum. Of particular interest, however, is the regime of strong light-matter coupling that was shown recently to drastically modify dispersion of electrons. To proper address this issue the system of electron and strong electromagnetic field has to be treated as a whole to form a bound electron-field state, referred to as Òelectron dressed by fieldÓ (dressed electron) \cite{Cohen-Tannoudji_b98,Scully_b01}. The field-induced modification of energy spectrum of dressed electrons has been previously studied in both atomic systems and various condensed-matter structures, including bulk semiconductors \cite{Elesin_69,Vu_04,Vu_05}, graphene \cite{Kibis_10,Kibis_11_1,Glazov_14,Usaj_14}, quantum wells \cite{Mysyrovich_86,Wagner_10,Teich_13,Kibis2015a}, quantum rings \cite{Kibis_13,Joibari_14}, and TIs \cite{Fregoso_2013,Calvo_2015}. Nevertheless, an ability to easily control and manipulate spin transport in field-dressed TIs, which is of great interest not only from fundamental viewpoint but also as a promising tool for future spintronics application, needs further investigation. In an attempt to fill the gap we examine a simple theoretical model which allows to include all relevant interactions and show that coupling between electrons and a linearly polarized electromagnetic wave leads to the anisotropy of electron energy spectrum near the Dirac point of TI. As a result, the strong light-induced modification of spin transport in TI takes place. Below we describe the details of our calculations.

\section{Theoretical model}

\begin{figure}[htb!]
\centering
\includegraphics[scale=0.445]{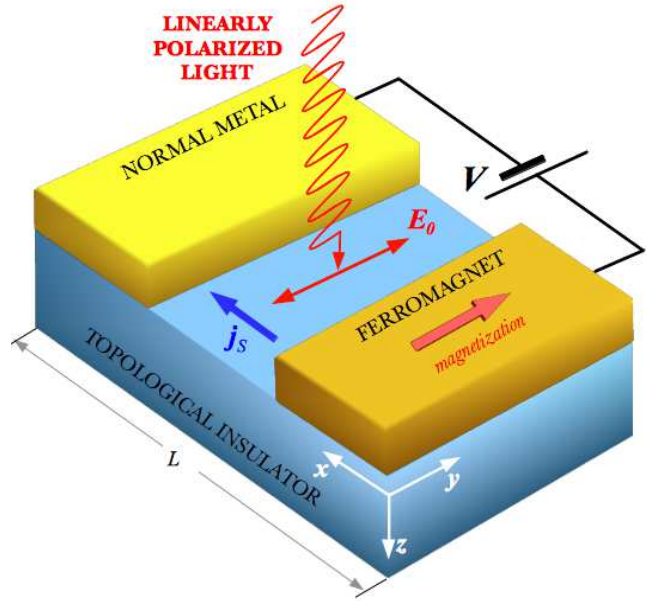}
\caption{\label{fig:figure1} The proposed setup is illustrated schematically: The surface of a 3D topological insulator irradiated by a linearly polarized electromagnetic wave with the electric field amplitude $E_0$. The magnetized ferromagnetic electrode injects the spin polarized current, $j_s$, which is fixed by the applied voltage $V$.}
\end{figure}

We consider electrons residing on the surface of TI and irradiated with a linearly polarized electromagnetic wave of the frequency $\omega$ and the electric field amplitude $E_0$, propagating perpendicularly to the surface (see Fig.~\ref{fig:figure1}). In the following, we will be to assume that the field frequency, $\omega$, is far from all characteristic resonances of the electron system, including the plasmon frequency (i.e., off-resonant or dressing field). The latter allows to neglect the processes accompanied by the energy exchange between the electron system and the field. The Hamiltonian of TI in the presence of the field is defined by \cite{Kane,Zhang},

\begin{equation}\label{H}
\hat{\mathcal{H}}=v_0\left[\bm{\sigma}\times\left(\hbar\mathbf{k}-\frac{e}{c}\mathbf{A}(t)\right)\right]_z,
\end{equation}

\noindent where $v_0$ is the electron velocity at the Dirac point of TI ($\mathbf{k}=0$), $e$ is the electron charge, $\bm{\sigma}=(\sigma_x,\sigma_y,\sigma_z)$ is the vector of Pauli matrices, $\mathbf{k}=(k_x,k_y)$ is the electron wave vector on the surface of TI, and $\mathbf{A}(t)=(c\mathbf{E}_0/{\omega})\cos\omega t$ is the vector potential of the wave which is assumed to be polarized along the $y$ axis. Formally, the Hamiltonian (\ref{H}) can be rewritten as $\hat{\mathcal{H}}=\hat{\mathcal{H}}_0+\hat{\mathcal{H}}_\mathbf{k}$, where $\hat{\mathcal{H}}_0=-\sigma_x(ev_0E_0/\omega)\cos\omega t$ is the Hamiltonian of dressed TI at the Dirac point, and $\hat{\mathcal{H}}_\mathbf{k}=\hbar v_0\left[\bm{\sigma}\times\mathbf{k}\right]_z$ is the Hamiltonian of the free quasiparticles on the surface of TI which are known to obey the Dirac dispersion. Solving the nonstationary Schr\"odinger equation, $i\hbar\partial\psi^\pm_0/\partial t=\hat{\mathcal{H}}_0\psi^\pm_0$, we easily arrive at the two wave functions of the dressed TI in the Dirac point,

\begin{equation}\label{p0}
\psi_0^\pm=\frac{1}{\sqrt{2}} \left(s_+\pm s_-\right)e^{\pm
i(\alpha/2)\sin\omega t},
\end{equation}

\noindent where $\alpha=(2ev_0E_0)/(\hbar\omega^2)$ is the parameter of electron-field coupling in TI, and $s_\pm$ are the spinors with the defined $z$ component of electron spin, $\sigma_zs_\pm=\pm s_\pm$. It should be stressed that the two wave functions (\ref{p0}) are exact: They describe a dressed electron in the Dirac point accurately. Therefore, from the beginning we treat an electromagnetic field in a non-perturbative manner. Since the wave functions (\ref{p0}) form the complete orthogonal system of the dressed TI at any given point in time $t$, we can find the nonstationary momentum-dependent $\mathbf{k}\neq0$ solution as an expansion

\begin{equation}\label{pk}
\psi_{\mathbf{k}}=a^+_\mathbf{k}(t)\psi^+_0+a^-_\mathbf{k}(t)\psi^-_0.
\end{equation}

\noindent Substituting the expansion (\ref{pk}) into the Schr\"odinger equation with the full Hamiltonian (\ref{H}), $i\hbar\partial\psi_{\mathbf{k}}/\partial t=\hat{\mathcal{H}}\psi_{\mathbf{k}}$, we derive a set of differential equations describing evolution of electrons in the dressed TI,

\begin{equation}\label{evolution}
i\dot{a}^\pm_\mathbf{k}(t)=\pm
v_0\left[k_ya^\pm_\mathbf{k}(t)-ik_xe^{\mp i\alpha\sin\omega
t}a^\mp_\mathbf{k}(t)\right].
\end{equation}

\subsection{Effective time-independent Hamiltonian}

Standard Floquet analysis suggests the wave function (\ref{pk}) to be of the form $\Psi(\mathbf{r},t)=e^{-i\tilde{\varepsilon}_\mathbf{k} t/\hbar}\phi(\mathbf{r},t)$, where the function $\phi(\mathbf{r},t)$ is time-periodic, $\phi(\mathbf{r},t)=\phi(\mathbf{r},t+2\pi/\omega)$, and $\tilde{\varepsilon}_\mathbf{k}$ is the Floquet quasi-energy of an electron (the energy of dressed electron) \cite{Hanggi1998,Aguado,Hanggi2005}. Fourier-expanding the coefficients $a^\pm_{\mathbf{k}}(t)$ in Eqs.~(\ref{pk}) and (\ref{evolution})

\begin{equation}\label{aF}
a^\pm_{\mathbf{k}}(t)=e^{-i\tilde{\varepsilon}_\mathbf{k}
t/\hbar}\sum_{n=-\infty}^{\infty}a^\pm_{n}({\mathbf{k}})e^{in\omega
t},
\end{equation}

\noindent substituting the expansion (\ref{aF}) into Eq.~(\ref{evolution}), and applying the Jacobi-Anger expansion, $e^{iz\sin\theta}=\sum_{n=-\infty}^{\infty}J_n(z)e^{in\theta}$ to transform the exponent in the right hand side of Eq.~(\ref{evolution}), one can rewrite the Schr\"odinger equation (\ref{evolution}) in the stationary form

\begin{equation}\label{Hk}
\sum_{n^\prime=-\infty}^{\infty}{\cal
H}_{nn^\prime}(\mathbf{k})a_{n^\prime}({\mathbf{k}})=
\tilde{\varepsilon}_\mathbf{k}a_{n}({\mathbf{k}}),
\end{equation}

\noindent where

\begin{equation}\label{Hk0}
{\cal H}_{nn^\prime}(\mathbf{k})=
\left(\begin{array}{cc}
\hbar\left(n\omega+v_0k_y\right)\delta_{nn^\prime} & -i\hbar
v_0k_x
J_{n^\prime-n}\left(\alpha\right) \\
i \hbar v_0k_xJ_{n^\prime-n}\left(\alpha\right) &
\hbar\left(n\omega-v_0k_y\right)\delta_{nn^\prime}
\end{array}\right)
\end{equation}

\noindent is the time-independent Floquet Hamiltonian of the dressed TI, $J_n(z)$ is the $n-$th order Bessel function of the first kind, $\delta_{mn}$ is the Kronecker symbol, and $a_{n}({\mathbf{k}})$ is the pseudospinor. It has to be said that the eigenvalue problem (\ref{Hk}) describes exactly the initial time-dependent problem in the extended Hilbert space.

In what follows, we assume

\begin{equation}\label{hf}
\left|\frac{J^2_{m}\left(\alpha\right)v_0k_x }
{{J_{0}\left(\alpha\right) }(2v_0k_y+m\omega)}\right|\ll1
\end{equation}

\noindent is satisfied with $m=\pm1,\pm2,\pm3...$. Mathematically, the expression (\ref{hf}) allows to consider off-diagonal matrix elements of the Hamiltonian (\ref{Hk0}) with $n\neq n^\prime$ as a small perturbation. Thus, the matrix elements can be omitted in the first-order perturbation theory. Physically, the constraint (\ref{hf}) results from the assumption that the field is off-resonant: The multi-photon energy, $\pm m\hbar\omega$, is far from resonant electron energies, $2\hbar v_0k_y$, corresponding to the electron transitions between different edge states of TI. Therefore, the field can be neither absorbed nor emitted by the electrons. As a consequence, the main contribution to the Schr\"odinger equation (\ref{Hk}) under the constraint (\ref{hf}) stems from terms with $n,n^\prime=0$, which describe the elastic interaction between an electron and the field. Within this approach, the Schr\"odinger problem for the dressed TI is reduced to the equation

\begin{equation}\label{Hkk}
{\cal H}_{00}(\mathbf{k})a_{0}({\mathbf{k}})=
\tilde{\varepsilon}_\mathbf{k}a_{0}({\mathbf{k}})
\end{equation}

\noindent with the effective Hamiltonian ${\cal H}_{00}(\mathbf{k})$ represented by the $2\times2$ matrix (\ref{Hk0}) with $n,n^\prime=0$. The effective Schr\"odinger equation (\ref{Hkk}) yields two linear algebraic equations for the two components of the spinor $a_{0}({\mathbf{k}})=(a_{0}^+({\mathbf{k}}),a_{0}^-({\mathbf{k}}))^T$. Working out these two equations gives rise to the energy spectrum of the dressed TI,

\begin{equation}\label{Ee}
\tilde{\varepsilon}^{(\pm)}_\mathbf{k}=\pm\hbar
v_0\sqrt{k_y^2+k_x^2J_0^2\left(\alpha\right)},
\end{equation}

\noindent and the corresponding eigenstates of the effective Floquet Hamiltonian,

\begin{equation}\label{a0}
a^{(\pm)}_{0}({\mathbf{k}})=\frac{e^{-i\tilde{\varepsilon}^{(\pm)}_\mathbf{k}t/\hbar}}{2}
\left(\begin{array}{c} \sqrt{\frac{\left|\tilde{\varepsilon}^{(\pm)}_\mathbf{k}\right|\pm\hbar
v_0k_y}{\left|\tilde{\varepsilon}^{(\pm)}_\mathbf{k}\right|}}\\
\pm
i\sqrt{\frac{\left|\tilde{\varepsilon}^{(\pm)}_\mathbf{k}\right|\mp\hbar
v_0k_y}{\left|\tilde{\varepsilon}^{(\pm)}_\mathbf{k}\right|}}\end{array}\right).
\end{equation}

\noindent The spinor (\ref{a0}) is written in the basis (\ref{p0}) and its upper and lower components are the coefficients $a_{0}^+({\mathbf{k}})$ and $a_{0}^-({\mathbf{k}})$, respectively. In the absence of external field ($\alpha=0$) the dispersion relation (\ref{Ee}) coincides with that of free electrons in TI \cite{Kane,Zhang}, $\varepsilon^{(\pm)}_{\bm{k}}=\pm\hbar v_0k$. Interestingly, the energy spectrum (\ref{Ee}) and the eigenstates (\ref{a0}) clearly manifest anisotropy in the $\mathbf{k}$ space owing to the Bessel function, $J_0(\alpha)$. Shining the system with a linearly polarized wave brings in a certain anisotropy: The isotropic round-shaped Fermi surface transforms into highly anisotropic ellipse, resulting in Fermi velocity along $x$ and $y$ directions differ. 

Besides the modification of electron energy spectrum, the dressing field renormalizes the scattering of conduction electrons in various low-dimensional systems \cite{Kibis2015a,Kristinsson_16}. To describe this field-induced effect in TI, we will apply the known scattering theory for dressed electrons in Dirac materials with linear electronic dispersion \cite{Kristinsson_16}. Restricting to the disorder constituted by randomly distributed delta-function scatterers \cite{Rammer,Burkov2}, one can write the scattering time for electrons at the Fermi energy of dressed TI as

\begin{equation}\label{tau}
\tilde{\tau}_F={\tau}_F\sqrt{|J_0(\alpha)|},
\end{equation}

\noindent where ${\tau}_F$ is the scattering time of free electrons in TI (see \ref{sec:appendix} for details).

\subsection{Spin injection from a ferromagnet to dressed TI}

In 3D topological insulators the surface transport is provided by a single mode with Dirac-like conical dispersion. Such a mode is an analog of unidirectional (chiral) mode at the edge of 2D system. The regime of strong light-matter coupling has so far completely avoided attention despite its potential importance in the emerging spinoptronics. The main focus of this section is, therefore, to extend the physical understanding developed previously for spin-injection from a ferromagnet to TI \cite{Mondal2010a,Mondal2010b,Palop2012,Ma2012,Kondo2014,Guo2014} to the case of dressed TI. 

We come with a proposal to make use of a setup shown in Fig.~\ref{fig:figure1}: We assume the system of a ferromagnet and a normal metal layers is placed on top of TI and the current flows parallel to the $x$ axis. If this is the case then in the weak disorder regime the only relevant interaction in the system is the proximity field due to ferromagnet polarized along the $y$ axis, such that the Hamiltonian ${\cal H}={\cal H}_{00}+Jm_y\sigma_y$, where $J$ is the exchange coupling and $m_y$ is the magnetization. Thanks to the translational symmetry in the transverse direction we can put $k_y=0$ in the vicinity of the neutral point. Assuming the depletion layer is narrow enough $U\rightarrow\infty$ and $L\rightarrow 0$, but keeping the dimensionless barrier strength $Z=UL$ finite, we can write down the corresponding wave functions: At $x<0$ occupied by a ferromagnet

\begin{equation}\label{psi1}
\psi_1(x\leq0)=\frac{e^{ik_xx}}{E\sqrt{2}}\left(\begin{array}{c}
v_xk_x-iJm_y \\ E
\end{array}\right)
+\frac{re^{-ik_xx}}{E\sqrt{2}}\left(\begin{array}{c}
-v_xk_x-iJm_y \\ E
\end{array}\right),
\end{equation}

\noindent TI with the Hamiltonian (\ref{Hkk}) at $0\leq x\leq L$

\begin{equation}\label{psi2}
\psi_2(0\leq x\leq L)=ae^{ik_x^{\prime\prime}x}\left(\begin{array}{c}
v_xk_x^{\prime\prime} \\ E-U
\end{array}\right)
+be^{-ik_x^{\prime\prime}x}\left(\begin{array}{c}
-v_xk_x^{\prime\prime} \\ E-U
\end{array}\right),
\end{equation}

\noindent and a normal metal at $x\geq L$

\begin{equation}\label{psi3}
\psi_3(x\geq L)=\frac{te^{ik_x^\prime x}}{E^\prime\sqrt{2}}\left(\begin{array}{c}
v_xk_x^\prime \\ E^\prime
\end{array}\right).
\end{equation}

\noindent In the formulas (\ref{psi1}), (\ref{psi2}), and (\ref{psi3}) the velocity $v_x=v_0$ for free electrons in TI, whereas $v_x=v_0\vert J_0(\alpha)\vert$ for dressed electrons. The coefficients $r$ and $t$ determine reflection and transmission amplitudes, while $a$ and $b$ are some constants to be identified from boundary conditions and $E$ denotes the dispersion relation of the Hamiltonian ${\cal H}$. The energy conservation leads to $E=U-v_xk_x^{\prime\prime}$ and $E^\prime=E-eV=v_xk_x^\prime$, where the potential drop $V$ can be tuned by the gate electrode. Matching the boundary conditions at $x=0$ and $x=L$: $\psi_1(0)=\psi_2(0)$ and $\psi_2(L)=\psi_3(L)$, we derive the transmission probability at the Fermi level $E=\tilde{\varepsilon}_F$

\begin{equation}\label{tr}
T=\vert t\vert^2=\frac{2\left(\tilde{\varepsilon}_F^2-J^2m_y^2\right)}{\tilde{\varepsilon}_F\left(\tilde{\varepsilon}_F+\sqrt{\tilde{\varepsilon}_F^2-J^2m_y^2}\right)}.
\end{equation}

\noindent Remarkably, one can see from the formula (\ref{tr}) that for $k_y=0$ the transmission probability is independent of barrier strength $Z$ and is purely determined by the energy of incident particles. For dressed electrons characterized by the dispersion relation (\ref{Ee}) the Fermi energy is defined as follows

\begin{equation}
\tilde{\varepsilon}_F=2\hbar v_0\sqrt{\pi n\vert J_0(\alpha)\vert}
\end{equation}

\noindent for a given concentration of charge carriers $n$. The current density is known to be proportional to the transmission amplitude, and as a consequence $\tilde{j}_s/j_s=\tilde{T}/T$, where $\tilde{j}_s\;(j_s)$ and $\tilde{T}\;(T)$ stand for dressed (free) electrons respectively. In full agreement with previous studies \cite{Mondal2010a,Mondal2010b} there exists a critical value of $m_y$ beyond which the current along $x$ direction becomes evanescent.

\subsection{Spin dynamics} 

In order to analyze the spin dynamics of the dressed TI under the influence of scattering processes, we adopt the conventional formalism based on the density matrix response function \cite{Rammer,Burkov2,Mahan,Burkov,Mischenko,Stanescu}. The comprehensive reviews of this theoretical technique can be found, for instance, in Refs.~\cite{Akkermans, Gumbs, Lerner}. Neglecting the interference of electron waves scattered by different scatterers, the evolution of the electron spin and charge can be described by the diffusion equation, ${\cal{D}}^{-1}\mathbf{S}=0$, where ${\cal{D}}$ is the spin-charge diffusion propagator (diffuson), and the four-vector $\mathbf{S}=(N,S_x,S_y,S_z)$ is introduced above to describe both the charge density, $N$, and the components of the spin density, $S_{x,y,z}$. Assuming the scattering processes in TI to be caused by a short-range ``white noise'' disorder, we can apply the model of delta-function scatterers which is relevant to describe the quantum transport in TIs \cite{Burkov2}. Then the diffuson in TI can be easily calculated by applying the diagram technique \cite{Burkov,Stanescu} to the system in question. Writing the diffuson as a sum of single joint scattering events diagrams,

\begin{equation}\label{diffuson}
{I}_{ij}=\frac{\hbar}{2\pi\tilde{\tau}_F\nu_F}
\int\limits_{\mathbf{k}_F}\frac{\mathrm{d}^2\mathbf{k}}{(2\pi)^2}\mathrm{Tr}\left[G^-_\mathbf{k}(\tilde{\varepsilon}_F)
\sigma_iG^+_\mathbf{k+q}(\tilde{\varepsilon}_F+\hbar\Omega)\sigma_j\right],
\end{equation}

\noindent we arrive at the expression ${\cal{D}}=(1-I_{ij})^{-1}$, where the frequency $\Omega$ and the wave vector $\mathbf{q}$ describe an external weak field (probing field) applied to TI, $\tilde{\varepsilon}_F=\tilde{\varepsilon}_{\mathbf{k}_F}$ is the Fermi energy of TI, $\nu_F$ is the density of states at the Fermi energy, $\tilde{\tau}_F$ is the impurity scattering rate at the Fermi energy, and $i,j=0,x,y,z$. Correspondingly, the symbols $G_\mathbf{k}^{\pm}$ in Eq.~(\ref{diffuson}) denote the disorder-averaged single-particle retarded (advanced) Green's functions which read as
\begin{equation}\label{green}
G^\pm_\mathbf{k}(\varepsilon)=\left[\varepsilon\pm i\hbar/(2\tilde{\tau}_F)-{\cal
H}_{00}(\mathbf{k})\right]^{-1}.
\end{equation}

\noindent Assuming the broadening to be small ($\tilde{\varepsilon}_F\tilde{\tau}_F\gg\hbar$), and the probing field to be smooth (${q}/{k}_F\ll1$) and slow ($\hbar\Omega/\tilde{\varepsilon}_F\ll1$), the solving of the diffusion equation, ${\cal{D}}^{-1}\mathbf{S}=0$, yields the set of coupled spin-charge dynamical equations in the dressed TI,

\begin{eqnarray}\nonumber
\frac{\partial S_x}{\partial
t}=-\frac{S_x}{{\tau}_F\sqrt{|J_0\left(\alpha\right)|}}+\frac{D}{2}|J_0\left(\alpha\right)|^{3/2}\frac{\partial^2S_x}{\partial
x^2}+\frac{3D\sqrt{|J_0\left(\alpha\right)|}}{2}\frac{\partial^2S_x}{\partial
y^2} \\ \label{x-component}
+D|J_0\left(\alpha\right)|^{3/2}\frac{\partial^2S_y}{\partial
x\partial y}-\frac{v_0}{2}\frac{\partial N}{\partial y},
\end{eqnarray}

\begin{eqnarray}\nonumber
\frac{\partial S_y}{\partial
t}=-\frac{S_y}{\tau_F\sqrt{|J_0\left(\alpha\right)|}}+\frac{3D}{2}|J_0\left(\alpha\right)|^{3/2}\frac{\partial^2S_y}{\partial
x^2}+\frac{D\sqrt{|J_0\left(\alpha\right)|}}{2}\frac{\partial^2S_y}{\partial
y^2} \\ \label{y-component}
+D|J_0\left(\alpha\right)|^{3/2}\frac{\partial^2S_x}{\partial
x\partial y}+\frac{v_0}{2}|J_0\left(\alpha\right)|\frac{\partial
N}{\partial x},
\end{eqnarray}

\noindent and

\begin{equation}\label{charge}
\frac{\partial N}{\partial
t}=D|J_0\left(\alpha\right)|^{3/2}\frac{\partial^2N}{\partial
x^2}+D\sqrt{|J_0\left(\alpha\right)|}\frac{\partial^2N}{\partial
y^2}-v_0\frac{\partial S_x}{\partial y}+v_0
|J_0\left(\alpha\right)|\frac{\partial S_y}{\partial x},
\end{equation}

\noindent where $D=v_0^2\tau_F/2$ is the diffusion coefficient of free electrons in TI. As expected, Eqs.~(\ref{x-component})--(\ref{charge}) exactly coincide with the known spin-charge dynamics equations for free electrons in TI \cite{Burkov2} when the dressing field is absent ($\alpha=0$).

\section{Discussion and conclusions}

\begin{figure}[htb!]
\centering
\includegraphics[scale=0.45]{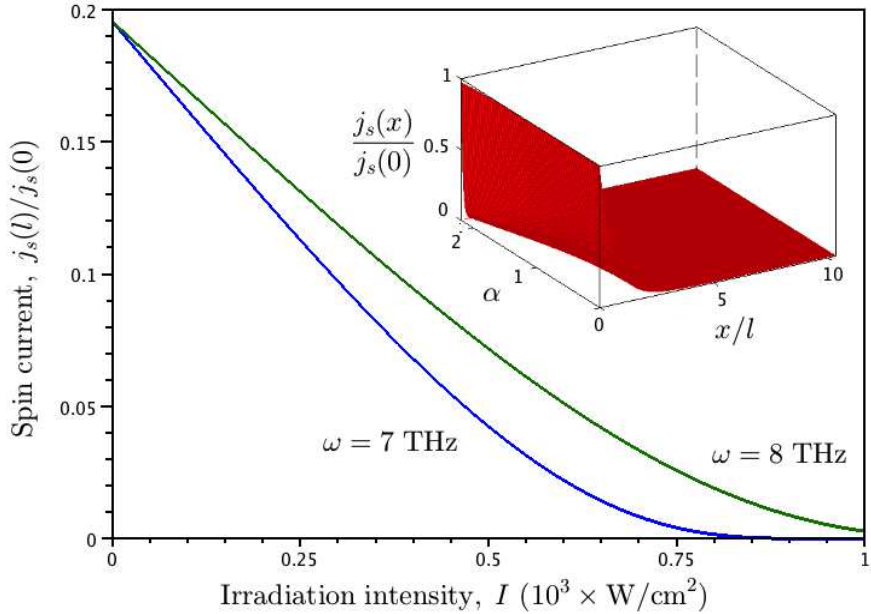}
\caption{\label{fig:figure2} Dependence of the spin current on the surface of a 3D topological insulator, $j_s$, on the irradiation intensity, $I$, for the two different frequencies of the irradiation, $\omega$. The inset shows the dependence of the spin current, $j_s$, on the electron-field coupling parameter, $\alpha$, and the coordinate $x$. In the calculations, the electron velocity at the Dirac point is assumed to be $v_0=5\cdot10^5$~m/s \cite{velocity}, the spin splitting $\Delta=0.16$, and the distance between the voltage electrodes is chosen to be $L=10l$, where $l=v_0\tau_F$ is the mean free path of electrons in the absence of the field.}
\end{figure}

To explore the spin effects emerging from the dressing field, we elaborate on the stationary spin current, $j_s$, induced by a stationary voltage, $V$, applied between electrodes on the surface of TI (see Fig.~\ref{fig:figure1}). Due to the last term in the right hand side of Eqs.~(\ref{x-component})--(\ref{charge}), the spin and charge densities are coupled to each other. As a result, an electric current between the electrodes is accompanied by the injection of spin polarized electrons from the ferromagnetic electrode. In the geometry depicted in Fig.~\ref{fig:figure1}, the electron spins are polarized along the $y$ axis and the spin current, $j_s$, flows parallel to the $x$ axis. Solving dynamical equations (\ref{x-component})--(\ref{charge}) equipped with the continuity equation and keeping in mind transmission probability (\ref{tr}), we derive for the spin current,

\begin{eqnarray}\nonumber
\frac{j_s(x)}{j_s(0)}=\frac{1}{\sqrt{\vert J_0(\alpha)\vert}}\frac{\vert J_0(\alpha)\vert-\Delta^2}{1-\Delta^2}\times \\ \label{js}
\times \frac{1+\sqrt{1-\Delta^2}}{\sqrt{\vert J_0(\alpha)\vert}+\sqrt{\vert J_0(\alpha)\vert-\Delta^2}} \frac{\sinh\left[{2\sqrt{2/3}(L-x)}/{l}\right]}{\sinh\left[{2\sqrt{2/3}L}/{l}\right]},
\end{eqnarray}

\noindent where $l={v}_0\tau_F|J_0(\alpha)|^{3/2}$, $L$ is the distance between the electrodes, and spin splitting $\Delta=Jm_y/\tilde{\varepsilon}_F$. It is worth noting that we assume  that an electromagnetic wave irradiating TI cannot be absorbed by electrons near the Dirac point. In the presence of scatterers, an off-resonance electromagnetic wave can be considered as a purely dressing (nonabsorbable) field if the wave frequency, $\omega$, meets $\omega\tau_F\gg1$. Physically, the latter means that the collisional absorption of the wave is negligibly small. Since the electron life time in TIs is $\tau_F\sim10^{-12}$~s \cite{time}, the field frequency should be high enough, $\omega>10^{12}$~Hz.  The effect of the high-frequency dressing field on the spin current is clearly visible in Fig.~\ref{fig:figure2}.

One can see in Fig.~\ref{fig:figure2} that the spin current, $j_s$, dramatically decreases with increasing the electron-field coupling parameter, $\alpha=(2ev_0E_0)/(\hbar\omega^2)$. Thus, the dressing field can strongly suppress the spin current. The physical reason of this light-induced spin effect is the following. Due to the giant spin-orbit coupling in TIs, the Elliott-Yafet mechanism of the spin relaxation \cite{Elliott,Yafet} is dominant for realistic scattering times of electrons, $\tau_F$. Since the scattering of spin-polarized electrons destroys the spin polarization, the spin current injected by the ferromagnetic electrode, $j_s(x)$, is appreciable only within the distance, $x\sim l$, where $l$ is the mean free path of dressed electrons. Close to the Dirac point of TI, the mean free path can be written as $\tilde{l}=|\tilde{v}_x|\tilde{\tau}_F$, where $\tilde{v}_x$ is the velocity of dressed electron along the $x$ axis, which can be derived from the energy spectrum (\ref{Ee}),

\begin{equation}\label{v0}
\tilde{v}_x=\left.\frac{1}{\hbar}\frac{\partial\tilde{\varepsilon}^{(\pm)}_\mathbf{k}}{\partial
k_x}\right|_{\mathbf{k}=0}=\pm v_0|J_0(\alpha)|.
\end{equation}

\noindent Therefore, the quantity $l={v}_0\tau_F|J_0(\alpha)|^{3/2}$ in Eq.~(\ref{js}) should be treated as a mean free path renormalized by the dressing field. Since $|J_0(\alpha)|\leq1$, the dressing field reduces both the electron velocity (\ref{v0}) and the mean free path of electrons, $l$. As a result, the light-induced suppression of the spin current, $j_s$, takes place.

According to the mentioned above, the discussed spin effects stem from the strong dependence of the electron velocity at the Dirac point (\ref{v0}) on the linearly polarized dressing field. It should be emphasized that the linear polarization of the field is crucial. In fact, illuminating Dirac materials with a circularly polarized light opens up a gap in the spectrum \cite{Kibis_10,Wang}, nevertheless its influence on the Fermi velocity is not so pronounced. Therefore, a linearly polarized dressing field seems most attractive from experimental viewpoint to observe the spin effects.

To summarize, we have worked out the theory of spin dynamics on the surface of a topological insulator (TI) in the regime of strong light-matter coupling. The theory reveals that the light-spin coupling drastically modifies the spin transport properties of TI. Particularly, it is shown that the irradiation of TI by an off-resonant high-frequency electromagnetic wave strongly suppresses the spin current. This effect can form a physical basis for optically tunable spintronic devices based on TIs. Of great interest is the optically controlled spin transistor. We expect that the predicted light-induced spin effects will motivate novel interdisciplinary activity at the border between spintronics \cite{Zutic,Spintronics} and optoelectronics since light-tunable electronic devices are typically much faster than those of electrically controlled.

\section*{Acknowledgements}

This work was partially supported by FP7 IRSES projects POLATER and QOCaN, Rannis project BOFEHYSS, RFBR projects 14-02-00033 and 16-32-60040, Singapore Ministry of Education under AcRF Tier 2 grant MOE2015-T2-1-055, Russian Ministry of Education and Science, and the Russian Target Federal Program ``Research and Development in Priority Areas of Development of the Russian Scientific and Technological Complex for 2014--2020'' (project RFMEFI58715X0020).

\appendix
\section{Derivation of the scattering time for dressed electrons}\label{sec:appendix}

In Dirac materials with linear electronic dispersion, the scattering probability of dressed electrons is described by Eqs.~(8)--(10) in Ref.~\cite{Kristinsson_16}. In particular, for topological insulators (TIs) illuminated with a linearly polarized electromagnetic wave, these expressions read
\begin{equation}\label{W1}
w_{\mathbf{k}^\prime\mathbf{k}}=\frac{2\pi}{\hbar}
|\chi_{\mathbf{k}^\prime\mathbf{k}}|^2\left|U_{\mathbf{k}^\prime\mathbf{k}}\right|^2
\delta(\tilde{\varepsilon}_{{\mathbf{k}}^\prime}-\tilde{\varepsilon}_{\mathbf{k}})
\end{equation}
and
\begin{equation}\label{chi_l}
\chi_{\mathbf{k}^\prime\mathbf{k}}=\sqrt{\frac{\cos\theta'+f(\theta')}{2f(\theta')}\frac{\cos\theta+f(\theta)}{2f(\theta)}}
\Bigg[1+\frac{J_0\left(\alpha\right)\sin\theta'}{\cos\theta'+f(\theta')}
\frac{J_0\left(\alpha\right)\sin\theta}{\cos\theta+f(\theta)}
\Bigg],
\end{equation}
where $w_{\mathbf{k}^\prime\mathbf{k}}$ is the probability of electron scattering per unit time from the initial electron state with the wave vector $\mathbf{k}=(k\sin\theta,k\cos\theta)$ to the final electron state with the wave vector $\mathbf{k}^\prime=(k^\prime\sin\theta^\prime,k^\prime\cos\theta^\prime)$, $\theta$ and $\theta^\prime$ are the corresponding azimuth angles, $U_{\mathbf{k}^\prime\mathbf{k}}$ is the matrix element of the disorder potential $U(\mathbf{r})$, and
\begin{equation}\nonumber
f(\theta)=\sqrt{\cos^2\theta+J_0^2\left(\alpha\right)\sin^2\theta}.
\end{equation}
Thus, the scattering time for the electron state $\mathbf{k}$ in dressed TI can be written as
\begin{equation}\label{s1}
\frac{1}{\tilde{\tau}(\mathbf{k})}=\sum\limits_{\mathbf{k}^\prime}w_{\mathbf{k}^\prime\mathbf{k}}=\frac{2\pi}{\hbar}\sum\limits_{\mathbf{k}^\prime}\vert\chi_\mathbf{k^\prime
k}\vert^2 \vert U_\mathbf{k^\prime
k}\vert^2\delta\left(\tilde{\varepsilon}_\mathbf{k^\prime}-\tilde{\varepsilon}_\mathbf{k}\right).
\end{equation}
For the disorder potential constituted by a succession of randomly distributed delta-function scatterers,
\begin{equation}\label{s3}
U(\mathbf{r})=\sum\limits_{i=1}^Nu_0\delta\left(\mathbf{r}-\mathbf{r}_i\right),
\end{equation}
we can write the squared scattering matrix element as
\begin{equation}\label{s5}
\left| U_{\mathbf{k}^\prime\mathbf{k}}\right|^2=\left|\frac{1}{S}
\int_Se^{i(\mathbf{k}-\mathbf{k}^\prime)\mathbf{r}}U(\mathbf{r})\mathrm{d}^2\mathbf{r}\right|^2=\frac{N_su_0^2}{S},
\end{equation}
where $S$ is the surface area of TI, and $N_s=N/S$ is the density of scatterers. Substituting Eq.~(\ref{s5}) into Eq.~(\ref{s1}), we can replace summation over the wave vector $\mathbf{k}^\prime$ with integration over $k^\prime$ and $\theta^\prime$,
$$\sum\limits_{\mathbf{k}^\prime}\rightarrow\int\frac{S\mathrm{d}^2\mathbf{k}^\prime}{(2\pi)^2}
\rightarrow\frac{S}{(2\pi)^2}\int
k^\prime\mathrm{d}k^\prime\int\mathrm{d}\theta^\prime,$$ which can be easily done. As a result, the scattering time (\ref{s1}) for electron states at the Fermi energy of dressed TI,
$\tilde{\varepsilon}_F$, reads
\begin{equation}\label{s6}
\frac{1}{\tilde{\tau}_F}=\frac{N_su_0^2\tilde{\varepsilon}_F}{2v_0^2\hbar^3}\frac{1}{|J_0(\alpha)|}.
\end{equation}
Taking into account that the dressing field does not change the density of charge carriers,
\begin{equation}\label{n}
n(\alpha)=\int_0^{\mathbf{k}_F}\frac{\mathrm{d}^2\mathbf{k}}{(2\pi)^2}=\frac{\tilde{\varepsilon}_F^2}{4\pi\hbar^2v_0^2}\frac{1}{|J_0(\alpha)|},
\end{equation}
we arrive at the equality $n(\alpha)=n(0)$. It follows from this that the Fermi energy in dressed TI, $\tilde{\varepsilon}_F$, acquires an extra field-dependent factor $\sqrt{|J_0(\alpha)|}$:
\begin{equation}\label{F}
\tilde{\varepsilon}_F={\varepsilon}_F\sqrt{|J_0(\alpha)|},
\end{equation}
where ${\varepsilon}_F$ is the Fermi energy of free electrons in TI. Substituting Eq.~(\ref{F}) into Eq.~(\ref{s6}), we derive for the renormalized scattering time in dressed TI,
\begin{equation}\label{tau1}
\tilde{\tau}_F={\tau}_F\sqrt{|J_0(\alpha)|},
\end{equation}
where $\tau_F=2v_0^2\hbar^3/N_su_0^2{\varepsilon}_F$ is the scattering time of free electrons in TI. As expected,
$\tilde{\tau}_F={\tau}_F$ in the absence of the dressing field
($\alpha=0$).

\end{document}